\documentclass[11pt]{iopart}

\usepackage{iopams}
\usepackage{amsfonts}
\usepackage{amsbsy} 
\usepackage{epsfig}
\usepackage{bbm}
\usepackage{latexsym}
\usepackage{showlabels}
\usepackage{color}
\usepackage[breaklinks]{hyperref}
\usepackage{tensor}
\newcommand{\ou}[3]{\tensor{#1}{^{#2}_{#3}}}
\usepackage{accents}
\usepackage{nicefrac}
\newcommand{\tfrac}[2]{\nicefrac{#1}{#2}}
\newcommand{\ubar}[1]{\underaccent{\bar}{#1}}
\usepackage{psfrag}
\usepackage[export]{adjustbox}
\newcommand{\oarX}[1]{\href{http://arxiv.org/abs/#1}{{\ttfamily #1}}}
\newcommand{\arX}[1]{\href{http://arxiv.org/abs/#1}{{\ttfamily arXiv:#1}}}

\def\ben{\begin{equation}}
\def\een{\end{equation}}
\def\bena{\begin{eqnarray}}
\def\eena{\end{eqnarray}}

\def\bR{\mathbb{R}}
\def\bC{\mathbb{C}}
\def\dd{\mathrm{d}}

\def\im{{\rm i}}

\begin{document}

\begin{flushright}\end{flushright}

\title{Cosmological evolution as squeezing: a toy model for group field cosmology}

\author{Eugene Adjei$^1$, Steffen Gielen$^{1,2,3,4}$, Wolfgang Wieland$^1$}
\address{$^1$Perimeter Institute for Theoretical Physics,\newline$\phantom{^1}$31 Caroline St.\ N., Waterloo, Ontario N2L\,2Y5, Canada}\vspace{0.2em}
\address{$^2$Max-Planck-Institut f\"ur Gravitationsphysik (Albert-Einstein-Institut),\newline$\phantom{^1}$Am M\"uhlenberg 1, 14476 Potsdam-Golm, Germany}\vspace{0.2em}
\address{$^3$Canadian Institute for Theoretical Astrophysics (CITA),\newline$\phantom{^1}$60 St.\ George Street, Toronto, Ontario M5S\,3H8, Canada}
\address{$^4$School of Mathematical Sciences, University of Nottingham,\newline$\phantom{^1}$University Park, Nottingham NG7 2RD, United Kingdom}
\ead{eadjei@perimeterinstitute.ca, sgielen@cita.utoronto.ca, wwieland@perimeterinstitute.ca}

\begin{abstract}
We present a simple model of quantum cosmology based on the group field theory (GFT) approach to quantum gravity. The model is formulated on a subspace of the GFT Fock space for the quanta of geometry, with a fixed volume per quantum.  In this Hilbert space, cosmological expansion corresponds to the generation of new quanta. Our main insight is that the evolution of a flat FLRW universe with a massless scalar field can be described on this Hilbert space as squeezing, familiar from quantum optics. As in GFT cosmology, we find that the three-volume satisfies an effective Friedmann equation similar to the one of loop quantum cosmology, connecting the classical contracting and expanding solutions by a quantum bounce. The only free parameter in the model is identified with Newton's constant. We also comment on the possible topological interpretation of our squeezed states. This paper can serve as an introduction into the main ideas of GFT cosmology without requiring the full GFT formalism; our results can also motivate new developments in GFT and its cosmological application.
\end{abstract}


\maketitle

\section{Introduction}

Cosmology is one of the main possibilities for how quantum gravity could become relevant for observations. Different approaches to quantum gravity give different scenarios for the fate of the initial singularity and for early universe cosmology, each with potentially different observational signatures \cite{Aurelien}. Loop quantum gravity (LQG), for instance, has led to loop quantum cosmology (LQC) \cite{LQC}, whose insights count among the main achievements of LQG: the classical singularity is resolved by a bounce, which connects a previous contracting to an expanding universe. In the framework of improved dynamics \cite{improdyn}, the late-time, semiclassical limit of LQC reduces to classical cosmology, with only a single new parameter introduced: a maximal (critical) energy density, which is of the order of the Planck density. If we go back in time until we reach this critical energy density, the fundamental quantum discreteness of space kicks in and prevents the Universe from contracting further. This provides an intuitive picture for how quantum effects can resolve cosmological singularities, see e.g.\ \cite{abhayIntro} for an introductory account. More recently, LQC has made contact with inflation, replacing the classical spacetime on which inflation is formulated by a so-called \emph{quantum spacetime} \cite{LQCinfl}.

One of the main questions concerning the foundations of LQC is its relation to the full theory of LQG. While some aspects of LQG, such as the discreteness of area and volume or the use of a polymer-like quantisation, are crucially used in the construction of LQC, there is as yet no fully satisfactory derivation of such models from LQG. What needs to be shown, in particular, is how the dynamics of LQC can emerge from some proposed dynamics of LQG, such as a Hamiltonian constraint in the canonical approach, or a spin foam model in the covariant formulation. This problem has attracted great interest recently, for example in the setting of quantum reduced loop gravity (QRLG) in which various features of LQC could be reproduced from a canonical formalism \cite{QRLG}. One major challenge faced by QRLG and related approaches (see, e.g., \cite{bodendorfer}), which to a large extent motivates the model we develop in this paper, is to justify the main assumption of LQC about the nature of quantum geometry: the assumption that the fundamental excitations that make up the Universe consist of certain minimal quanta, such that the expansion of the Universe corresponds to creation of new quanta rather than inflating existing ones. In canonical LQG, this would presumably require constructing a new, graph-changing Hamiltonian that can generate new quanta (spin network nodes) while preserving an appropriate notion of homogeneity.

As is well-known, the description of quantum systems in which the number of quanta changes dynamically is often easiest in quantum field theory, where one has field operators that create and annihilate particles. This is precisely what we will do in this paper to develop a toy model for a quantum description of cosmology. Incorporating the main idea of LQC, we will assume that each quantum of geometry comes with a fixed (Planckian) volume. We then show how the cosmological dynamics for a free, massless scalar field in a flat Friedmann--Lema\^itre--Robertson--Walker (FLRW) universe corresponds classically to a dilatation of volume and quantum-mechanically to squeezing, familiar from quantum optics. By studying the expectation value of the total three-volume relative to the scalar field, we show that the quantum dynamics defined by a squeezing Hamiltonian is in agreement with the classical theory for late times. The resulting effective equations are very similar to the LQC effective equations, where the classical contracting and expanding solutions are connected by a non-singular bounce. An appealing feature of squeezing is its preservation of uncertainty relations for elementary phase space variables: squeezing of a state (e.g.\ the Fock vacuum) of minimal uncertainty results in a highly excited, coherent state that also saturates the Heisenberg bound for these variables. We interpret this property as the emergence of a large, semiclassical universe, described by a large number of quanta with respect to the vacuum, following precisely the classical Friedmann dynamics at low curvature.

Our model fits well into the effective cosmological models developed within the group field theory (GFT) approach over the last years \cite{GFTreview}. GFT provides a second quantised language for LQG, with field operators that create and annihilate quanta of geometry. The key idea is then that a macroscopic cosmological universe should correspond to a GFT {\em condensate}, a coherent quantum configuration of a large number of quanta. Using methods from the study of Bose\,--\,Einstein condensates in condensed matter physics, one can derive effective equations for such GFT condensates that can be interpreted in cosmological terms. Adding a massless scalar field as matter, one can link these equations to those of LQC. It could be shown, in particular, that GFT condensates undergo a bounce; moreover, assuming that all quanta in the condensate have the same microscopic volume (as in LQC), effective Friedmann equations could be derived, very similar to those of LQC \cite{GFTfried}. The last assumption can be further motivated by showing explicitly, in a wide class of GFT models, that for a more generic initial condensate state a single component (corresponding to a single volume eigenvalue) will always dominate asymptotically \cite{lowspin}.

Our model is constructed to reproduce the classical dynamics of an FLRW universe. It is not derived from any proposed GFT action. On the other hand, we do not need to make assumptions about the emergence of a condensate phase; {\em any} initial state will result in a large universe following the classical Friedmann dynamics. Thus, our model provides a proof of principle that the full physical evolution of quantum geometry states can lead to states of condensate type, and that one can connect to classical cosmology and to LQC starting from a simple discrete model of quantum geometry. 

The insights gained from our analysis could become useful for developments in GFT. In particular, an important difference between our model and usual GFT concerns the choice of canonical commutator algebra for the field operators. Taking the role of the massless scalar field as a relational clock seriously, we propose {\em equal-time} commutation relations, which is not what is usually done in GFT where no fundamental notion of time is used. Similarly, the role of squeezing as cosmological time evolution might suggest possible dynamics for full GFT which reproduce cosmological dynamics in a more direct way. At the end of the paper, we also comment on the possible topological interpretation of our squeezed states.

\section{FLRW cosmology with a scalar field}
\label{cosmosec}

We consider a flat FLRW universe filled with a free, massless and homogeneous scalar field as matter. This is a very simple cosmological model, whose dynamics can be deparametrised by using the scalar as a clock \cite{blythisham}, effectively treating its value $\phi$ as a time variable (this is justified because $\frac{\dd\phi}{\dd t}$ never changes sign and so $\phi$ evolves monotonically). 

In an FLRW universe the evolution of the spatial geometry is conventionally given in terms of the scale factor $a(t)$ such that the physical three-metric is $h_{ij}(t)=a^2(t)h^{0}_{ij}$ in terms of a fixed `fiducial' (here flat) metric $h^{0}_{ij}$. The phase space variables for gravity and matter are then $a$ and $\phi$ with their canonical momenta $p_a$ and $\pi_\phi$. There is a single constraint corresponding to the freedom of time reparametrisations, given by the Friedmann equation (see, e.g., \cite{bojobook})
\ben
\mathcal{C} = -\frac{2\pi G}{3}\frac{p_a^2}{a} + \frac{1}{2}\frac{\pi_\phi^2}{a^3} = 0\,.
\een
In the usual Dirac formalism for constrained Hamiltonian systems, one would impose $\mathcal{C}=0$ as a constraint and define a Hamiltonian $N\mathcal{C}$, where $N$ is the (arbitrary) lapse function, to generate dynamics.

Deparametrisation amounts to identifying a suitable degree of freedom, here the scalar $\phi$, as a time variable with respect to which `true' evolution can be defined. We then need to choose one of the square roots of the Friedmann equation, leading to a Hamiltonian
\ben
\pi_\phi = \mathcal{H}:=\pm\sqrt{\frac{4\pi G}{3}}\,a\, p_a\,.
\label{depfriedmann}
\een
After deparametrisation, the phase space variables are $a(\phi)$ and $p_a(\phi)$ which are unconstrained. This formalism can be the starting point for a quantisation in which (\ref{depfriedmann}) becomes the Schr\"odinger equation for a wavefunction of the Universe; this is indeed what is done in loop quantum cosmology (LQC) where the right-hand side is replaced by a suitable operator well-defined on the LQC Hilbert space ($p_a$, which involves a connection, is not; technically speaking, its exponential is not weakly continuous in the quantum theory \cite{abhayIntro}, hence $p_a$ does not exist as an operator itself).

Notice that for this cosmological model time evolution corresponds to a dilatation; the equations of motion are
\ben
\frac{\dd a}{\dd\phi}=\pm\sqrt{\frac{4\pi G}{3}}\,a\,,\quad \frac{\dd p_a}{\dd\phi}=\mp\sqrt{\frac{4\pi G}{3}}\, p_a
\een
and their solutions are obviously exponential in $\phi$, corresponding to an expanding or a contracting universe depending on the choice of sign.

Time evolution corresponds to a dilatation not only for $a$ but also for any power of $a$; if we pass from $a$ to the volume $V\sim a^3$, we have
\ben
\mathcal{H} = \pm\sqrt{12\pi G}\,Vp_V
\label{dephamiltonian}
\een 
and again exponential solutions (of course, any power of an exponential is also an exponential). The volume is the variable most commonly used in LQC and we will focus on it in the following.

In this classical deparametrised formalism, one needs to choose a sign in (\ref{depfriedmann}) leading to either only expanding or only contracting solutions. These approach a singularity as $\phi\rightarrow-\infty$ (Big Bang) or $\phi\rightarrow +\infty$ (Big Crunch), respectively. The achievement of LQC \cite{LQC} was to define a quantum evolution that interpolates between these classical alternatives, and connects a contracting to an expanding universe through a non-singular quantum bounce. We will see something similar in our model: for large positive (negative) values of $\phi$ the evolution of the Universe is well approximated by the flow of the classical Hamiltonian (\ref{dephamiltonian}) for a positive (negative) chosen sign, while a deviation is found from the classical theory at high curvature (near $\phi=0$).

\section{Quantum cosmology as squeezing}
\label{qcsqueeze}

We could now set up a ``first quantised'' quantum theory in which $\mathcal{H}$ in (\ref{dephamiltonian}) becomes the Hamiltonian acting on a wavefunction $\psi(V,\phi)$. The resulting Schr\"odinger equation can be derived in the usual way from an action for $\psi$ and its complex conjugate $\bar\psi$, 
\ben
S[\psi,\bar\psi]=\int \dd\phi \int \dd V\left[\frac{\im}{2}\left(\bar\psi\frac{\dd\psi}{\dd\phi}-\psi\frac{\dd\bar\psi}{\dd\phi}\right)-\bar\psi\hat{\mathcal{H}}\psi\right]
\label{action1}
\een
where $\hat{\mathcal{H}}$ is an appropriate Hermitian operator representing the quantum Hamiltonian, e.g.\ $\hat{\mathcal{H}}=\sqrt{3\pi G}(\hat{V}\hat{p}_V+\hat{p}_V\hat{V})$. 

Notice that, instead of the classical phase space variables $V(\phi)$ and $p_V(\phi)$, (\ref{action1}) defines dynamics for a field $\psi(V,\phi)$ and its complex conjugate. The first term in the action plays the role of the symplectic form $p\,dq$, showing that $\psi$ and $\bar\psi$ are canonically conjugate, and the second term introduces a field Hamiltonian $\mathcal{H}_\psi := \int\dd V\bar\psi\hat{\mathcal{H}}\psi$ defining the dynamics. This action viewpoint on Schr\"odinger quantum mechanics provides an immediate starting point for `second quantisation' in which one now views $\psi$ and $\bar\psi$ as field operators in a quantum field theory, with dynamics defined by the action (\ref{action1}) or its extension to an interacting theory in which terms of higher order in $\bar\psi$ or $\psi$ can be added to $\mathcal{H}_\psi$.

We can then adopt such a second-quantised viewpoint on quantum cosmology in which, rather than defined in terms of a Schr\"odinger-type (single-particle) wavefunction, the state of the Universe is made up of many elementary quantum patches or `geometric atoms' governed by a quantum field theory. This viewpoint has been advocated from various directions including quantum cosmology \cite{martinreview}, and is in line with the insights obtained over the last decades in loop quantum gravity and related approaches such as GFT: geometry is itself quantised at the Planck scale, and a macroscopic, homogeneous universe should really arise from the interactions of a very large number of such quanta of geometry. A second quantised approach also provides a direct route to including inhomogeneities, which can correspond to a slightly inhomogeneous many-particle configuration or, somewhat similar to inflation, directly arise as fluctuations in the quantum field that generates geometry \cite{GFTinhomo}.

A simple possibility would be to promote (\ref{action1}) directly to the action of a quantum field theory; we would then have field operators $\hat\Psi$ and $\hat\Psi^\dagger$ with canonical commutation relations (here treating $V$ as a real variable that can also take negative values)
\ben
[\hat\Psi(V,\phi),\hat\Psi^\dagger(V',\phi)]=\delta(V-V')
\label{qccommut}
\een
and the Hamiltonian would be $\int\dd V\hat\Psi^\dagger\hat{\mathcal{H}}\hat\Psi$ with some differential operator $\hat{\mathcal{H}}$. This field theory would be non-interacting, with dynamical equations that are linear in the fields. In particular, the dynamics would conserve the particle number $\int \dd V \langle \hat\Psi^\dagger \hat\Psi \rangle$, just like the norm of a wavefunction is conserved in single-particle quantum mechanics.
 
This possibility has rather undesirable consequences for cosmology; it would suggest that the number of geometric quanta has remained constant while the total volume of the Universe has increased by many orders of magnitude. As in this scenario expansion could only proceed by expansion of the quanta themselves, initially Planck-size quanta would be macroscopic today. Not only do we not have any evidence for discreteness in the Universe around us, but these large quanta would presumably not be able to support short enough wavelengths for cosmological perturbations (known as the trans-Planckian problem in inflation \cite{transplanck}). Moreover, the improved dynamics prescription for LQC \cite{improdyn} suggests that expansion of the Universe proceeds purely through generation of new quanta of geometry, where these quanta remain at constant (Planckian) volumes at all times. Connecting in any way to LQC requires us to change the dynamics such that particle number is not conserved. Recent results in GFT condensates have already shown how, similar to LQC, the expansion of the Universe can be understood as generation of new quanta of geometry \cite{GFTfried,lowspin}. All this motivates us to define a different type of dynamics for cosmology.

A related point is that, if there are indeed fundamental quanta of geometry, an approximately continuous macroscopic universe must consist of many quanta in a highly symmetric configuration, in order to correspond to the great simplicity (homogeneity, isotropy) of the observed Universe on largest scales. This has motivated the idea of describing the Universe as a kind of {\em condensate}, a macroscopic coherent configuration of many quanta. In the context of GFT, condensates have been the main tool to connect to cosmology; a condensate is defined by the property that a quantum state of many quanta is fixed by a single-particle wavefunction \cite{GFTreview}. In particular, `dipole condensate' states have appeared in this context \cite{condJHEP} that are very similar to squeezed states in quantum optics. Compared to the simpler mean-field coherent states, dipole condensates have the advantage of being naturally gauge-invariant from the perspective of LQG, and so possessing a clearer geometric interpretation (see section~\ref{toposec} for a discussion of their topological interpretation in GFT).

In this paper, we propose a simple model for GFT cosmology that combines these insights with the fact that, for a massless scalar field in a flat universe, time evolution in $\phi$ corresponds to exponential expansion of the spatial volume. We will make use of well-known properties of squeezed states in quantum optics \cite{quop}: consider a single harmonic oscillator, with associated creation and annihilation operators $\hat{a}^\dagger$ and $\hat{a}$. Starting from the Fock vacuum $|0\rangle$, one can define a squeezed state as
\ben
|\zeta\rangle = \hat{S}(\zeta)|0\rangle := \exp\left(\frac{\zeta}{2}\hat{a}^\dagger\hat{a}^\dagger-\frac{\bar{\zeta}}{2}\hat{a}\hat{a}\right)|0\rangle\,.
\label{squeezedstate}
\een
The action of $\hat{S}(\zeta)$ corresponds to a Bogoliubov transformation, i.e.\ a change of basis of creation and annihilation operators; by the Baker--Campbell--Hausdorff formula,
\bena
\hat{S}^\dagger(\zeta)\hat{a}\hat{S}(\zeta) &=& \hat{a} + \zeta\hat{a}^\dagger + \frac{1}{2}|\zeta|^2\hat{a} + \frac{1}{6}|\zeta|^2 \zeta\hat{a}^\dagger + \ldots \nonumber
\\&=& \cosh(|\zeta|)\hat{a} + \sinh(|\zeta|)\frac{\zeta}{|\zeta|}\hat{a}^\dagger\,.
\eena
It follows that, with respect to the original Fock vacuum, the number of quanta in the squeezed state $|\zeta\rangle$ is
\ben
\langle \zeta|\hat{a}^\dagger\hat{a}|\zeta\rangle = \langle 0|\hat{S}^\dagger(\zeta)\hat{a}^\dagger\hat{S}(\zeta)\hat{S}^\dagger(\zeta)\hat{a}\hat{S}(\zeta)|0\rangle = \sinh^2(|\zeta|)\,.
\een
For large $|\zeta|$, this grows exponentially in $|\zeta|$. Squeezing thus realises exactly the exponential growth in the particle number needed for cosmology. 

One can see directly that the squeezing operator effectively acts as a dilatation in the particle number $n$, at least in the limit where the latter is large: if we take $\zeta$ to be real, squeezing corresponds to the exponentiated action of a Hermitian operator $\hat{s}$,
\ben
\hat{S}(\zeta)=\exp(-\im \zeta\hat{s})\,,\quad \hat{s}=\frac{\im}{2}\left(\hat{a}^\dagger\hat{a}^\dagger-\hat{a}\hat{a}\right)\,.
\een
The action of this operator $\hat{s}$ on a normalised particle number eigenstate $|n\rangle\equiv(n!)^{-1/2}(\hat{a}^\dagger)^n |0\rangle$ is
\bena
\hat{s}|n\rangle &=& \frac{\im}{2\sqrt{n!}}\left((\hat{a}^\dagger)^{n+2}-n(n-1)(\hat{a}^\dagger)^{n-2}\right)|0\rangle\nonumber
\\ &=&\frac{\im}{2}\left(\sqrt{(n+1)(n+2)}|n+2\rangle-\sqrt{n(n-1)}|n-2\rangle\right)\,.
\eena
For large $n$ and in a continuum limit, $\hat{s}$ acts just like a dilatation in $n$, $\hat{s}\sim -2\im (n\frac{\partial}{\partial n}+\frac{1}{2})$ (we will derive the numerical factors in more detail below). In the following we will develop a cosmological toy model for GFT that implements the main insight of the improved dynamics prescription for LQC, namely that the total volume is proportional to the number of quanta. Squeezing then not only acts as dilatation in the particle number but also in the cosmological volume, as suggested by the classical Friedmann dynamics (\ref{dephamiltonian}).

Another interesting property of squeezing is the preservation of certain uncertainty relations. Consider the elementary Hermitian operators $\hat{a}+\hat{a}^\dagger$ and $\im(\hat{a}-\hat{a}^\dagger)$, which would correspond to position and momentum for a harmonic oscillator. Under squeezing, these transform as
\ben
\hat{S}^\dagger(\zeta)(\hat{a}+\hat{a}^\dagger)\hat{S}(\zeta) = e^\zeta (\hat{a}+\hat{a}^\dagger)\,,\quad \hat{S}^\dagger(\zeta)\im(\hat{a}-\hat{a}^\dagger)\hat{S}(\zeta) = \im e^{-\zeta}(\hat{a}-\hat{a}^\dagger)\,;
\een
expectation values and all higher moments of these operators are hence simply rescaled by squeezing. The product of fluctuations $\Delta(\hat{a}+\hat{a}^\dagger)\Delta(\im(\hat{a}-\hat{a}^\dagger))$ is conserved, and equal to its minimal value (unity) for the state (\ref{squeezedstate}) for all $\zeta$. In this sense, squeezed states have semiclassical properties rather similar to those of GFT condensates that describe macroscopic geometries.

These observations lead us to our main proposal: {\em cosmological time evolution is best realised as squeezing}. 

\section{Toy model for GFT cosmology: kinematics}

We will build on work of the last years on GFT condensates \cite{GFTreview} to develop a model for cosmology in which time evolution corresponds to squeezing an initial state (such as the Fock vacuum) to obtain a generalised `condensate'. GFT itself defines a second quantisation formalism for loop quantum gravity, i.e.\ a quantum field theory of geometry, in which creation and annihilation operators corresponding to quanta of geometry are defined naturally  \cite{GFT2nd}. 

In the cosmological context we are interested in, the starting point is a GFT for gravity coupled to a massless scalar field, in four spacetime dimensions. Here one usually starts with a complex bosonic field whose arguments are four elements of a Lie group $G$ and a real variable corresponding to the massless scalar,
\ben
\varphi:G^4\times \bR \rightarrow \bC\,.
\een
The group $G$ corresponds to the local gauge group of internal frame rotations. In the Ashtekar--Barbero formalism it would be $G=SU(2)$ which we choose here for definiteness. 

One imposes a symmetry under right multiplications of all four group elements,
\ben
\varphi(\ubar{g},\phi)\equiv\varphi(g_1,\ldots,g_4,\phi)=\varphi(g_1h,\ldots,g_4h,\phi)\quad\forall h\in SU(2)
\label{gaugeinv}
\een
corresponding to discrete gauge transformations in a sense that will become clear shortly.

Dynamics for a GFT is then usually defined either through a path integral, whose expansion into Feynman amplitudes corresponds to a sum over discrete spacetime histories, or through the canonical formalism as developed in \cite{condJHEP,GFT2nd}. In the latter, one introduces canonical commutation relations
\ben
\left[\hat\varphi(\ubar{g},\phi),\hat\varphi^\dagger(\ubar{g}',\phi')\right]=\delta(\phi-\phi')\int_{SU(2)}\!\!\!\!  \dd h\;\prod_{I=1}^4 \delta(g'_I h g_I^{-1})
\label{gftcommut}
\een
where $\ubar{g}=(g_1,\dots,g_4)$ and the integral (with respect to the normalised Haar measure $\dd h$)  over all $h\in SU(2)$ ensures that the commutation relations are compatible with the \emph{right invariance} of the GFT field operator  (\ref{gaugeinv}). Next, one introduces the Fock space starting from the vacuum $|\emptyset\rangle$, which is annihilated by the field operator: $\hat\varphi(\ubar{g},\phi)|\emptyset\rangle=0$, such that $\hat\varphi^\dagger$ creates an `atom of space' from $|\emptyset\rangle$. Schematically, we may write
\setlength{\unitlength}{0.030cm}
\begin{center}
\begin{picture}(320,100)
\put(7.5,50){$\hat\varphi^{\dagger}(g_1,g_2,g_3,g_4,\phi)\big|\emptyset\big\rangle=\Bigg|$}\put(290,50){$\Bigg\rangle$}
\put(222,52){$\bullet$}\put(220,100){\line(-1,-3){30}}\put(190,10){\line(3,1){90}}\put(280,40){\line(-1,1){60}}\put(190,10){\line(-1,1){30}}\put(160,40){\line(1,1){60}}
\put(165,60){$g_1$}\put(245,90){$g_2$}\put(245,40){$g_3$}\put(202,30){$g_4$}
\put(217,60){$\phi$}
\thicklines
\put(225,55){\line(-1,0){65}}\put(225,55){\line(1,1){45}}\put(225,55){\line(1,-1){50}}\put(225,55){\line(-1,-3){20}}
\end{picture}
\end{center}
and view the state $\hat\varphi^\dagger(\ubar{g},\phi)|\emptyset\rangle$ as a chunk of space, a tetrahedron whose geometrical degrees of freedom are characterised by four $SU(2)$-valued parallel transports through its four faces, and a label $\phi$ corresponding to the value of the scalar field. The four links meet at a central vertex where a discrete gauge transformation would indeed map $g_I\mapsto g_Ih$, as in (\ref{gaugeinv}). At this stage, the links are seen as open with no gauge transformations acting on the other end.

It is now convenient to use a Peter--Weyl decomposition of the GFT field into $SU(2)$ irreducible representations. Namely, one writes (see e.g.\ \cite{GFTfried}\footnote{Compared to \cite{GFTfried}, we have changed factors of $2j_I+1$ so that the terms under the product sign are normalised.})
\ben
\hat\varphi(\ubar{g},\phi)=\sum_{\ubar{j},\iota} \,\hat\varphi^{\ubar{m}}(\ubar{j},\iota,\phi)\,\mathcal{I}^\ast_{\ubar{n}}(\ubar{j},\iota) \prod_{I=1}^4 \sqrt{2j_I+1}\; \ou{D^{(j_I)}(g_I^{-1})}{n_I}{m_I}
\label{peterweyl}
\een
where $D^{(j_I)}(g_I)$ is the Wigner $D$-matrix for the $SU(2)$ element $g_I$ in the spin-$j_I$ representation and $\ubar{m}$ is the multi-index $\ubar{m}=(m_1,\dots,m_4)$ and equally $\ubar{j}=(j_1,\dots,j_4)$. The entries of $D^{(j_I)}(g_I)$  are labelled by magnetic indices $m_I$ and $n_I$. Using Einstein's summation convention, we sum over all repeated magnetic indices. The appearance of the $SU(2)$ invariant tensors\footnote{The defining property is: $\forall h\in SU(2):\ou{D^{(j_1)}(h)}{m_1}{n_1}\cdots\ou{D^{(j_4)}(h)}{m_4}{n_4}\mathcal{I}^{\ubar{n}}(\ubar{j},\iota)=\mathcal{I}^{\ubar{m}}(\ubar{j},\iota)$}  $\mathcal{I}^{\ubar{n}}(\ubar{j},\iota)\in j_1\otimes\dots\otimes j_4$  is a consequence of the right invariance property (\ref{gaugeinv}) of the GFT field operator. These \emph{intertwiners} are labelled by an index $\iota$, which runs over an orthonormal basis in the $SU(2)$ invariant (singlet) subspace of $j_1\otimes\dots\otimes j_4$, hence
\begin{equation}
\langle\mathcal{I}(\ubar{j},\iota),\mathcal{I}(\ubar{j},\iota')\rangle\equiv\mathcal{I}^\ast_{m_1\dots m_4}(\ubar{j},\iota)\mathcal{I}^{m_1\dots m_4}(\ubar{j},\iota')=\delta_{\iota\,\iota'}\,.
\end{equation}
All this implies now that the definition (\ref{peterweyl}) can be inverted for the coefficients $\hat\varphi^{\ubar{m}}(\ubar{j},\iota,\phi)$,
\ben
\hat\varphi^{\ubar{m}}(\ubar{j},\iota,\phi) = \int_{SU(2)^4}\!\!\!\!\dd^4 g\;\mathcal{I}^{\ubar{n}}(\ubar{j},\iota)\,\hat\varphi(\ubar{g},\phi)\prod_{I=1}^4\sqrt{2j_I+1}\,\ou{D^{(j_I)}(g_I)}{m_I}{n_I}\,.\label{peterweyl2}
\een
An analogous expression for the Hermitian conjugate field $\hat\varphi^\dagger$ in terms of Peter--Weyl modes ${\hat\varphi}^\dagger_{\ubar{m}}(\ubar{j},\iota,\phi)$ is obtained by taking the Hermitian conjugate of \eref{peterweyl}.

In the geometric interpretation of GFT states given by loop quantum gravity, the spins $j_I$ correspond to possible eigenvalues for the areas of the faces of the tetrahedron, which are given (in units of $\hbar=c=1)$ by $A_I=8\pi\gamma G\,\sqrt{j_I(j_I+1)}$, where $\gamma$ is the Barbero--Immirzi parameter and $G$ is Newton's constant. Thus, expressing the GFT field in a spin representation rather than the group representation means that we focus on metric information (areas) rather than connection information as given by parallel transports $g_I$.

The commutation relation (\ref{gftcommut}) implies that the field operators $\hat\varphi^{\ubar{m}}$ and ${\hat\varphi}^\dagger_{\ubar{m}}$ satisfy
\newpage
\bena
\Big[\hat\varphi^{\ubar{m}}(\ubar{j},&\iota,\phi),{\hat\varphi}^\dagger_{\ubar{m}'}(\ubar{j}',\iota',\phi')\Big]=\delta(\phi-\phi')\int_{SU(2)^4}\!\!\!\!\dd^4 g\;\mathcal{I}^{\ubar{r}}(\ubar{j},\iota)\,\mathcal{I}^\ast_{\ubar{s}}(\ubar{j}',\iota')\times\nonumber
\\&\qquad\times\prod_{I=1}^4 \sqrt{(2j_I+1)(2j_I'+1)}\;\ou{D^{(j_I)}(g_I)}{m_I}{r_I}\ou{D^{(j'_I)}(g_I^{-1})}{s_I}{m'_I}\nonumber
\\&=\delta(\phi-\phi')\delta_{\ubar{j}\ubar{j}'}\,\delta^{\ubar{m}}_{\ubar{m}'}\,\mathcal{I}^{\ubar{r}}(\ubar{j},\iota)\,\mathcal{I}^\ast_{\ubar{r}}(\ubar{j}',\iota')\nonumber
\\&= \delta(\phi-\phi')\,\delta_{\ubar{j}\ubar{j}'}\,\delta^{\ubar{m}}_{\ubar{m}'}\,\delta_{\iota\,\iota'}\,.
\eena
The group averaging over $h$ is taken care of by the intertwiners contracting the $r$ and $s$ indices, and in the second line we used $g^\dagger=g^{-1}$ and the orthogonality of the Wigner matrices with respect to the Haar measure $\dd g$.

These are the commutation relations of creation and annihilation operators. Notice that we have a delta function in $\phi$, inherited from (\ref{gftcommut}), so that `atoms' can be created independently at different values of $\phi$. In such a formalism the scalar field variable $\phi$ is just another direction in the configuration space of the GFT field; $\phi$ does not play the role of a physical time variable which would be suggested by the deparametrised cosmological formalism of section~\ref{cosmosec}.

In this paper, as we have argued, we are interested in a model in which $\phi$ plays the role of a time variable. Correspondingly, as in standard quantum field theory where a time variable is given by a background spacetime, we assume {\em equal-time} commutation relations. In the Heisenberg picture these would be of the form
\ben
\Big[\hat{a}^{\ubar{m}}(\ubar{j},\iota,\phi),\hat{a}^\dagger_{\ubar{m}'}(\ubar{j}',\iota',\phi')\Big]=\delta_{\ubar{j}\ubar{j}'}\,\delta^{\ubar{m}}_{\ubar{m}'}\,\delta_{\iota\,\iota'}\,,
\label{acommut}
\een
where we write $\hat{a}$ and $\hat{a}^\dagger$ to make clear that this formalism is different from the one derived from (\ref{gftcommut}). One can switch to a Schr\"odinger picture in which the operators have no $\phi$ dependence but states evolve in time, as we will do later on. In this sense, we are proposing a GFT toy model in which the field operators have the canonical commutation relations (\ref{qccommut}) of `second quantised quantum cosmology', but whose dynamics does not preserve particle number. The commutation relations (\ref{acommut}) will be derived from an action, unlike in the usual canonical formalism for GFT, where they are postulated, {\em a priori}.

The GFT dynamics involves all modes, i.e.~all possible values of $j_I$ and $\iota$, corresponding to all possible sizes and shapes of tetrahedra. To build a model for cosmology, we truncate the theory so that only some of the modes are excited. First of all, and following \cite{GFTfried}, we restrict ourselves to isotropic tetrahedra for which all spins are equal. This seems to be sufficient for building a macroscopic geometry which is itself isotropic. Then we follow the general expectation coming from LQC \cite{improdyn} that the relevant modes are those corresponding to minimal non-zero eigenvalues of the area, i.e.\ those for which $j=\tfrac{1}{2}$.\footnote{The role of $j=0$ quanta in GFT is somewhat different than in LQG; they are `soft tetrahedra' that have zero area or volume, but contribute to the total particle number. To facilitate comparison with LQG and LQC, we set the number of $j=0$ quanta to zero.}

In the GFT Peter--Weyl expansion (\ref{peterweyl}), we would focus only on the term
\ben
\hat\varphi(\ubar{g},\phi)=4\sum_{\iota=\iota^{\pm}}\hat\varphi^{A_1\dots A_4}(\iota,\phi)\mathcal{I}^\ast_{B_1\dots B_4}
\ou{[g^{-1}_1]}{B_1}{A_1}\cdots\ou{[g^{-1}_4]}{B_4}{A_4}\,,
\label{expansionterm}
\een
where we write $\ou{[g_I]}{A}{B}\equiv \ou{D^{(\frac{1}{2})}(g_I)}{A}{B}$ for the fundamental representation of $SU(2)$. Here, the magnetic indices $A,B,C,\dots$ correspond to spinor indices; we distinguish between `upstairs' and `downstairs' indices which correspond, respectively, to the fundamental representation and its complex conjugate; and indices are raised and lowered using the Hermitian metric $\psi^\ast_A=\delta_{AA'}\bar{\psi}^{A'}$. This distinction is conventional for $SU(2)$ spinor indices.

For all $j_I$ taken to be $\tfrac{1}{2}$, the space of intertwiners is two-dimensional. Two independent and orthogonal intertwiners, which may be denoted $\iota^+$ and $\iota^-$, correspond to the eigenvectors of an LQG operator corresponding to the oriented squared volume, with positive and negative eigenvalue given by (see \cite{bianchihaggard} for a summary of how to compute such eigenvalues in LQG)
\begin{equation}
\pm v_o^2=\pm\frac{(8\pi\gamma G)^3}{6\sqrt{3}}\,.\label{volval}
\end{equation}
For simplicity, we restrict ourselves to the intertwiner $\iota^+$ which corresponds to a positive orientation, and drop the intertwiner label in the following. In the usual canonical GFT formalism, we would then obtain a field operator $\hat\varphi^{ABCD}(\phi)$ and its Hermitian conjugate, with commutation relations
\ben
[\hat\varphi^{A_1\dots A_4}(\phi),\hat\varphi^\dagger_{B_1\dots B_4}(\phi')]=\delta(\phi-\phi')\delta^{A_1}_{B_1}\cdots\delta^{A_4}_{B_4}\,.
\een
In our formalism in which $\phi$ corresponds to time, we now instead introduce Schr\"odinger-picture operators $\hat{a}^{ABCD}$ and $\hat{a}^\dagger_{ABCD}$ with fundamental commutation relations
\ben
[\hat{a}^{A_1\dots A_4},\hat{a}^\dagger_{B_1\dots B_4}]=\delta^{A_1}_{B_1}\cdots\delta^{A_4}_{B_4}\,.
\een

We can then introduce a Fock space for these operators, starting from a vacuum $|0\rangle$ annihilated by all annihilation operators, $\hat{a}^{ABCD}|0\rangle=0$. The resulting Fock space for this GFT toy model can be seen as a subspace of a Fock space for a GFT based on equal-time commutation relations, with field operators satisfying
\ben
\left[\hat\Phi(\ubar{g},\phi),\hat\Phi^\dagger(\ubar{g}',\phi)\right]=\int \dd h\;\prod_{I=1}^4 \delta(g'_I h g_I^{-1})\,.
\een
The Fock space for $\hat{a}^{ABCD}$ and $\hat{a}^\dagger_{ABCD}$ includes those quanta within the larger Fock space for $\hat\Phi$ and $\hat\Phi^\dagger$ for which only the representation labels $j_I=\tfrac{1}{2}$ and the intertwiner $\iota^+$ are being excited. From the perspective of LQG, these are quanta with minimal non-zero area and volume, which are symmetric in the sense of describing equilateral chunks of geometry. LQC suggests using such quanta to build a cosmological universe.

We make one further simplification in the model. Namely, the tensor product of four fundamental representations can be decomposed into irreducible representations of $SU(2)$ according to
\ben
\frac{1}{2}\otimes\frac{1}{2}\otimes\frac{1}{2}\otimes\frac{1}{2} = 0 \oplus 1 \oplus 1 \oplus ( 0\oplus 1\oplus 2)\,.
\een
Concretely, the creation operators $\hat{a}^{ABCD}$ can be written as
\bena
\hat{a}^{ABCD} &=& \frac{1}{4}\epsilon^{AB}\epsilon^{CD}\hat{I}_{(1)}+\frac{1}{2}(\epsilon^{AC}\epsilon^{BD}+\epsilon^{BC}\epsilon^{AD})\hat{I}_{(2)}\nonumber\\
&&+\frac{1}{2}\epsilon^{AB}\hat{V}^{CD}_{(1)}+\frac{1}{2}\epsilon^{CD}\hat{V}^{AB}_{(2)}+\frac{1}{2}(\epsilon^{AC}\hat{V}^{BD}_{(3)}+\epsilon^{BD}\hat{V}^{AC}_{(3)})\nonumber
\\&&+\hat{A}^{ABCD}
\eena
where all operators on the right-hand side have totally symmetric indices. We now assume that only the totally symmetric component $\hat{A}^{ABCD}=\hat{A}^{(ABCD)}$, i.e.\ the spin-2 component of the tensor product of four fundamental representations, is excited. From the GFT perspective this would mean imposing an additional symmetry under permutations of the four arguments. In cosmology, we may use this simplification as an additional restriction to implement isotropy. We then use only five out of 16 oscillator modes given by the $\hat{a}^{ABCD}$.

\section{Toy model for GFT cosmology: dynamics}

We now define the dynamics for our model. Classically, the dynamical variables are the totally symmetric oscillator modes $A^{ABCD}$ and their complex conjugates. We take the action to be of the form
\ben
S[A^i,{A}^\ast_i]=\int \dd\phi\left[\frac{\im}{2}\left({A}^\ast_i\frac{\dd A^i}{\dd\phi}-\frac{\dd A^\ast_i}{\dd\phi}A^i\right)-\mathcal{H}(A^i,A^{\ast}_i)\right]
\label{action2}
\een
in close analogy to (\ref{action1}). We write $A^i\equiv A^{A_1\dots A_4}$ where $i$ is a magnetic index running over the five totally symmetric combinations of four spinor indices, and $A^\ast_i$ denotes the Hermitian conjugate with respect to the $SU(2)$ metric: $A^\ast_i\equiv A^\ast_{A_1\dots A_4}=\delta_{A_1B_1'}\dots\delta_{A_4B_4'}\bar{A}^{B_1'\dots B_4'}$. The Hamiltonian $\mathcal{H}$ is now chosen to violate particle number conservation, and in order to model cosmological time evolution we choose it to be a squeezing operator,
\ben
\mathcal{H} \equiv \frac{\im}{2}\lambda({A}^\ast_i{A}^\ast_j\epsilon^{ij} - A^i A^j\epsilon_{ij})\,.
\een
The $i$ and $j$ indices are contracted with appropriate combinations of the invariant tensor $\epsilon^{AB}=-\epsilon^{BA}$ for the spinor representation, i.e.\ $A^i A^j\epsilon_{ij}\equiv A^{A_1\dots A_4}A^{B_1\dots B_4}\epsilon_{A_1B_1}\dots\epsilon_{A_4B_4}$. The inverse $\epsilon$-tensor is given by $\epsilon^{ij}$, $\epsilon^{im}\epsilon_{jm}=\delta^i_j$. Notice also $\epsilon_{ij}=\epsilon_{ji}$.

The coupling constant $\lambda$ must be real for $\mathcal{H}$ to be real. In principle one could also introduce a complex coupling (and then multiply the second term by $\bar{\lambda}$). However, choosing $\lambda$ to be real is no loss of generality: the kinetic term in the action (\ref{action2}) is invariant under a (global) $U(1)$ transformation
\ben
A^i \rightarrow e^{\im\theta}A^i\,,\quad {A}^\ast_i \rightarrow e^{-\im\theta}{A}^\ast_i\,.
\een
Such a field redefinition, which does not alter the dynamical content of (\ref{action2}), sends $\lambda\rightarrow e^{-2\im\theta}\lambda$ and $\bar\lambda\rightarrow e^{2\im\theta}\bar\lambda$ in the more general case; thus $\lambda$ can always be made real by an appropriate phase transformation. 

As for quantum cosmology (\ref{action1}), the first term in (\ref{action2}) determines the Poisson brackets that turn $A^i$ and ${A}^\ast_i$ into canonically conjugate operators at the quantum level,
\ben
[\hat{A}^i,\hat{A}^\dagger_j]=\delta^i_j\,.
\een
We now have a choice of working in the Heisenberg picture or the Schr\"odinger picture. In the Schr\"odinger picture, we have a $\phi$-dependent Fock state that evolves according to the Schr\"odinger equation
\ben
\im\frac{\dd}{\dd\phi}|\chi(\phi)\rangle = \frac{\im}{2}\lambda(\hat{A}^\dagger_i \hat{A}^\dagger_j \epsilon^{ij} - \hat{A}^i \hat{A}^j\epsilon_{ij})|\chi(\phi)\rangle
\een
with general solution
\ben
|\chi(\phi)\rangle = \exp\left(\frac{\lambda}{2}\phi\left(\hat{A}^\dagger_i \hat{A}^\dagger_j \epsilon^{ij} - \hat{A}^i \hat{A}^j\epsilon_{ij}\right)\right)|\chi_o\rangle
\een
where $|\chi_o\rangle$ is an arbitrary initial state: physical states are obtained by acting with a $\phi$-dependent squeezing operator on any initial state. The assumption of GFT condensate cosmology \cite{GFTreview} that there are coherent condensate states that well approximate physical GFT states describing macroscopic, homogeneous geometries is thus realised explicitly within the context of a simplified model. Semiclassical properties of a squeezed state with respect to elementary operators suggest that in this model an asymptotically classical universe emerges dynamically from, e.g.~the Fock vacuum $|0\rangle$. The same squeezed states do generally not have small uncertainties in the volume, as one might demand for the emergence of a classical geometry (and as has been shown in the context of GFT condensates \cite{Pithis}); we will discuss volume uncertainties below. At the level of expectation values, the model we propose is able to reproduce Friedmann-like dynamics of a flat FLRW universe with a massless scalar field, as we will show next.

We consider only a single possible eigenvalue $v_o$ for the volume per tetrahedron, corresponding to the intertwiner $\iota^+$ (see discussion below (\ref{expansionterm}) and  \eref{volval}). In this reduced model, the GFT volume operator $\hat{V}$ is reduced to a multiple of the number operator,
\begin{equation}
\hat{V}=v_o\hat{A}^\dagger_i\hat{A}^i=v_o\hat{N}\,.
\end{equation}
Dilatation with respect to the volume, as in the classical cosmological Hamiltonian (\ref{dephamiltonian}), is then equivalent to `dilatation' with respect to the particle number (in an approximate sense, given that the latter is discrete), which is in turn realised by squeezing. We saw this already for a single harmonic oscillator. For five oscillator modes given by $\hat{A}^i$ and $\hat{A}^\dagger_j$, we can similarly define normalised states
\ben
|k\rangle = 2^{-(k+1)}\sqrt{\frac{3}{k!(k+\frac{3}{2})(k+\frac{1}{2})\cdots(\frac{1}{2})}}(\epsilon^{ij}\hat{A}^\dagger_i \hat{A}^\dagger_j)^k|0\rangle\,;
\een
the squeezing Hamiltonian then acts as
\bena
\hat{\mathcal{H}}|k\rangle&=&\frac{\im}{2}\lambda\left(\hat{A}^\dagger_i \hat{A}^\dagger_j \epsilon^{ij} - \hat{A}^i \hat{A}^j\epsilon_{ij}\right)|k\rangle\nonumber
\\&= &\frac{\im}{2}\lambda\left(\sqrt{(2k+2)(2k+5)}|k+1\rangle-\sqrt{2k(2k+3)}|k-1\rangle\right)\,.\label{Hactn}
\eena
For large $k$ and in the $v_o\rightarrow 0$ continuum limit this corresponds to the action of a dilatation operator. This can be seen as follows: Introduce a state $\Psi(V)\in L^2(\mathbb{R}_+,\dd V)$ in the continuum, and define its \emph{shadow state} on the lattice: $|\Psi\rangle:=\sum_{k=0}^\infty \Psi_k|k\rangle$, for components $\Psi_k=\sqrt{v_o}\Psi(k v_o)$ (the normalisation $\sqrt{v_o}$ is introduced such that for two such states $\Psi_k$ and $\Psi'_k$ the sum $\sum_{k=0}^\infty\bar{\Psi}_k\Psi_k'$ returns the $L^2$ inner product in the $v_o\rightarrow 0$ continuum limit). By duality, i.e.\ using $\langle k|\hat{\mathcal{H}}|\Psi\rangle=\overline{\langle\Psi|\hat{\mathcal{H}}|k\rangle}$, we now find the difference equation
\begin{eqnarray}
(\hat{\mathcal{H}}\Psi)(V)&=\frac{\lambda}{2\im v_o}\left(\sqrt{(2V+2v_o)(2V+5v_o)}\Psi(V+v_o)\right.\nonumber\\
&\hspace{8em}\left.-\sqrt{2V(2V+3v_o)}\Psi(V-v_o)\right)
\end{eqnarray}
for any $V=kv_o$. Assuming the first derivative $\partial_V\Psi(V)$ exists, we can now use L'H\^{o}pital's rule to take the continuum limit and find
\begin{eqnarray}\nonumber
(\hat{\mathcal{H}}\Psi)(V)\stackrel{v_o\rightarrow 0}{\rightarrow}-2\lambda\,\mathrm{i}\left(V\partial_V+\frac{1}{2}\right)\Psi(V)\,.
\end{eqnarray}
Setting $\lambda:=\sqrt{3\pi G}$ and using the symmetric ordering $\frac{1}{2}(\hat{V}\hat{p}_V+\hat{p}_V\hat{V})$ for the  product $Vp_V$, we thus recover the deparametrised Hamiltonian (\ref{dephamiltonian}) in the large-volume and continuum limit.

The time-evolution operator corresponding to our Hamiltonian,
\ben
\hat{S}(\lambda\phi):=\exp\left(\frac{\lambda}{2}\phi\left(\hat{A}^\dagger_i \hat{A}^\dagger_j \epsilon^{ij} - \hat{A}^i \hat{A}^j\epsilon_{ij}\right)\right)\,,
\een
also again realises a Bogoliubov transformation of the creation and annihilation operators, namely
\bena
\hat{S}^\dagger(\lambda\phi)\hat{A}^i\hat{S}(\lambda\phi) &=& \hat{A}^i + \lambda\phi\epsilon^{ij}\hat{A}_j^\dagger + \frac{1}{2}(\lambda\phi)^2\hat{A}^i + \frac{1}{3!}(\lambda\phi)^3\epsilon^{ij}\hat{A}_j^\dagger + \ldots\nonumber
\\& =& \cosh(\lambda\phi)\hat{A}^i + \sinh(\lambda\phi)\epsilon^{ij}\hat{A}_j^\dagger
\eena
and similarly for $\hat{A}^\dagger_i$. From this, the number of quanta in the state $|\chi(\phi)\rangle$ is found to be
\bena
N(\phi)&\equiv&\langle\chi(\phi)|\hat{A}_i^\dagger\hat{A}^i|\chi(\phi)\rangle\nonumber
\\ &=& \cosh^2(\lambda\phi)\langle \chi_o|\hat{A}_i^\dagger\hat{A}^i|\chi_o\rangle + \sinh^2(\lambda\phi)\langle \chi_o|\hat{A}^i\hat{A}_i^\dagger|\chi_o\rangle\nonumber
\\&&+\cosh(\lambda\phi)\sinh(\lambda\phi)\left(\epsilon^{ij}\langle \chi_o|\hat{A}_i^\dagger\hat{A}_j^\dagger|\chi_o\rangle +\epsilon_{ij}\langle \chi_o|\hat{A}^i\hat{A}^j|\chi_o\rangle\right)\nonumber
\\& =& -\frac{5}{2}+\Big(N_0+\frac{5}{2}\Big)\cosh(2\lambda\phi)  +\mathfrak{Re}(Q)\sinh(2\lambda\phi)\,,
\eena
where $N_0:=\langle \chi_o|\hat{A}_i^\dagger\hat{A}^i|\chi_o\rangle$ is the expectation value of the total particle number in the chosen initial state $|\chi_o\rangle$, and $Q:= \epsilon^{ij}\langle \chi_o|\hat{A}_i^\dagger\hat{A}_j^\dagger|\chi_o\rangle$. For simple initial states, for example eigenstates of the number operator, $Q=0$. In general, a nonzero $Q$ will render the bounce asymmetric in $\phi$. 
\\For this $\phi$-dependent total particle number, we then observe the following properties:
\begin{itemize}
\item At late or early times, $\phi\rightarrow\pm\infty$, the 3-volume $V(\phi)=v_oN(\phi)$ asymptotes to
\ben
V(\phi)=v_o\langle\chi(\phi)|\hat{A}_i^\dagger\hat{A}^i|\chi(\phi)\rangle \sim v_o\Big(\frac{N_0}{2} + \frac{5}{4}\pm\frac{\mathfrak{Re}(Q)}{2}\Big)\exp(2|\lambda\phi|)
\label{vasympt}
\een
with the sign given by the sign of $(\lambda\phi)$. The three-volume hence interpolates between a contracting and an expanding solution of the classical cosmological dynamics of section~\ref{cosmosec} if we fix $\lambda:=\sqrt{3\pi G}$. Newton's constant is `emergent' from the coupling constant $\lambda$ in our GFT toy model, in much the same way that it emerges from fundamental GFT couplings in \cite{GFTfried}.
\item Because $N(\phi)$ is always non-negative, a singularity where $N(\phi)=0$ is also a minimum at which $N'(\phi)=0$. These equations only have a solution for $\phi$ if
\ben
|\mathfrak{Re}(Q)| = \sqrt{N_0(N_0+5)}\,;
\label{specialvalue}
\een
a singularity can only be encountered for special initial conditions. It is also impossible for $N(\phi)$ to reach zero as $\phi\rightarrow\pm \infty$ (as in the classical theory); from (\ref{vasympt}), this would require $|\mathfrak{Re}(Q)| = N_0+\frac{5}{2}$, but this would in fact imply $N(\phi)\rightarrow -\frac{5}{2}$ asymptotically, which is impossible.
\\In summary, generic states in the Hilbert space avoid the classical singularity and undergo a bounce connecting the classical expanding and contracting branches. At the bounce, the three-volume reaches
its minimum, where $N(\phi)>0$. A singularity appears only for very special initial conditions (\ref{specialvalue}).
\end{itemize}
Let us assume $Q=0$ from now on; then time evolution is symmetric in $\phi$, as for the LQC effective dynamics (i.e.~for suitable states for which these are valid). In this case, we also see that:
\begin{itemize}
\item Unless the initial state is chosen to be the Fock vacuum $|0\rangle$, the number of quanta and the total three-volume are bounded away from zero: $N(\phi)\ge N_0$ at all times, with equality only at $\phi=0$. In this sense one finds a bounce resolving the classical singularity, again very similar to what was found for full GFT in \cite{GFTfried}. For the Fock vacuum, one obviously starts with zero particle number and hence a singularity in the geometric interpretation, but $\phi$-evolution still results in a large universe following exactly the classical Friedmann dynamics. As for a single harmonic oscillator, these states remain semiclassical in the sense that
\ben
\Delta(\hat{A}^i+\epsilon^{ij}\hat{A}_j^\dagger)\Delta(\im(\hat{A}^i-\epsilon^{ij}\hat{A}_j^\dagger))={\rm const.}
\een
For fluctuations of the volume $V$, we find that generically $\Delta V \sim V$ at late times, i.e.~the relative uncertainty $(\Delta V)/V$ approaches a constant. This behaviour is expected, since classically $\{\mathcal{H},V\}= \pm\sqrt{12\pi G}\,V$ and thus one would expect
\ben
\Delta\hat{\mathcal{H}}\,\Delta \hat{V}\geq \sqrt{3\pi G}\langle \hat{V}\rangle\,,
\een
where $\Delta\hat{\mathcal{H}}$ is a constant that just depends on the initial state. One would then look for states for which $(\Delta V)/V\ll 1$ at late times; see \cite{LQCuncertainties} for a discussion of the analogous issue in the context of LQC. In our model, simple initial states such as the Fock vacuum have $(\Delta V)/V \rightarrow O(1)$ at late times.
\item The volume satisfies the effective Friedmann equation
\ben
\left(\frac{1}{V}\frac{\dd V}{\dd\phi}\right)^2 = 4\lambda^2 \left(1+ \frac{5v_o}{V(\phi)}-\frac{N_0(N_0+5)v_o^2}{V(\phi)^2}\right)\,.
\label{efffried}
\een
Of the three terms in brackets, the first just gives the classical Friedmann equation (again, with $\lambda=\sqrt{3\pi G}$). The third term can be written as $\rho/\rho_c$, for some maximal (critical) energy density $\rho_c$, given that the energy density $\rho$ of the massless scalar field scales like $V^{-2}$; such a term appears in the effective Friedmann equations valid for suitable semiclassical states in LQC \cite{taveras}, and is responsible for the bounce. Here this term is absent if $N_0=0$, the case of the initial state chosen to be the Fock vacuum in which the singularity is not resolved. Indeed, for $N_0=0$ the effective Friedmann equation shows no repulsion at high density. The second term is another quantum correction, effectively behaving like an ultra-stiff (or ekpyrotic) matter component with equation of state $w=2$.

Structurally this effective Friedmann equation appears extremely similar to the one found for full GFT, for the case of isotropic (equilateral) GFT condensates in which only a single spin is excited, i.e.\ essentially the case we consider in our toy model. There \cite{GFTfried}, one finds
\ben
\left(\frac{1}{V}\frac{\dd V}{\dd\phi}\right)^2 = 12\pi G + \frac{4v_o E}{V(\phi)} - \frac{4v_o^2\pi_\phi^2}{V(\phi)^2}
\een
where $E$ (the GFT `energy') and $\pi_\phi$ (the scalar field momentum) are conserved, state-dependent quantities. For further discussion of such effective Friedmann equations in terms of effective matter components, for condensates with only a single excited $j$ but including different GFT interactions, see also \cite{mairietal}. 
\item The effective Friedmann equation (\ref{efffried}) assumes a particularly familiar form if we use a symmetric ordering for the GFT volume operator, which shifts the volume operator by a constant,
\begin{equation}
\hat{V}_{\rm s}=\frac{v_o}{2}(\hat{A}^\dagger_i \hat{A}^i+\hat{A}^i \hat{A}^\dagger_i)=v_o\Big(\hat{A}^\dagger_i \hat{A}^i+\frac{5}{2}\Big)\,.
\end{equation}
The expectation value $V_{\rm s}(\phi)=\langle\chi(\phi)|\hat{V}_{\rm s}|\chi(\phi)\rangle$ satisfies the effective Friedmann equation 
\begin{equation}
\left(\frac{1}{V_{\rm s}}\frac{\dd V_{\rm s}}{\dd\phi}\right)^2 = 4\lambda^2\left(1-\frac{v_o^2(N_0+\frac{5}{2})^2}{V_{\rm s}(\phi)^2}\right)\,.
\end{equation}
This is exactly of the LQC effective dynamics form
\ben
\left(\frac{1}{V}\frac{\dd V}{\dd\phi}\right)^2 = 12\pi G\left(1-\frac{\rho}{\rho_{\rm c}}\right)
\een
if we identify $\rho:=\lambda^2(N_0+\frac{5}{2})^2/V_{\rm s}^2$ with the energy density of matter, $\rho_{\rm c}:=\frac{\lambda^2}{v_o^2}$ with the (Planckian) critical density, and $\lambda=\sqrt{3\pi G}$.
\end{itemize}
Our toy model hence reproduces several of the results for effective Friedmann equations found in GFT condensate cosmology, without requiring the assumption of a condensate state: as we have shown, even starting from the Fock vacuum or a one-particle initial state leads to a squeezed `condensate' state whose dynamics can mimic LQC effective dynamics and resolve the classical singularity. For the Fock vacuum there is, by assumption, an initial geometric singularity in which no quanta were present, but the evolution is nevertheless regular. For the most general initial state (with $Q\neq 0$), we saw that almost all states still go through a bounce that avoids the classical singularity, and satisfy the classical Friedmann dynamics at early and late times. However, in general the bounce is asymmetric, and the connection to LQC and previous work on GFT condensate cosmology is less direct.

\section{Spatial topology and the GFT Fock space}
\label{toposec}

The construction of the Fock space for our model, and more generally for GFT, suggests a possibility of associating topological information to the states, in addition to geometric observables such as the three-volume we have discussed in the application to cosmology. Namely, if a single-particle state is pictured as a tetrahedron, contracting the indices associated to open links could be interpreted as `gluing', i.e.\ topological identification. A two-particle state such as
\begin{equation}
\includegraphics[scale=0.2,valign=c]{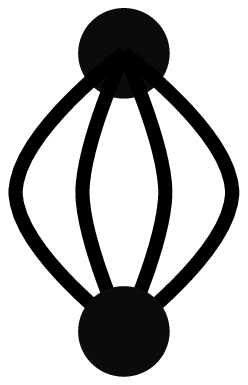}\big|0\rangle=\epsilon^{ij}A^\dagger_{i}A^\dagger_{j}|0\rangle
\end{equation}
could then be interpreted as the triangulation of a 3-sphere by two tetrahedra with all four faces identified. Such an interpretation is natural in simplicial geometry, and often also assumed in discussions of LQG spin network states. 

It would follow that applying the `dipole creation operator' $\hat{A}^\dagger_i \hat{A}^\dagger_j \epsilon^{ij}$ twice to obtain
\ben
\includegraphics[scale=0.2,valign=c]{dipole.eps}\,\includegraphics[scale=0.2,valign=c]{dipole.eps}\big|0\rangle=\left(\hat{A}^\dagger_i \hat{A}^\dagger_j \epsilon^{ij}\right)^2|0\rangle
\label{twodipoles}
\een
produces a state corresponding, topologically, to two disconnected three-spheres. Taking this interpretation further to define the topology of the geometries represented by our squeezed states would imply that, rather than representing a macroscopic cosmological universe, these squeezed states, and the condensate states of full GFT \cite{GFTreview}, correspond to a large number of disconnected, Planck-size universes, rendering their physical meaning unclear.

In the context of our toy model, it is easy to see that such an interpretation is not consistent as it is necessarily ambiguous. Take the state (\ref{twodipoles}), which is explicitly proportional to
\bena\nonumber
\hat{A}^\dagger_i \hat{A}^\dagger_j \hat{A}^\dagger_k \hat{A}^\dagger_l  \epsilon^{ij}\epsilon^{kl} |0\rangle &\equiv& \hat{A}^\dagger_{A_1A_2A_3A_4} \hat{A}^\dagger_{B_1B_2B_3B_4} \hat{A}^\dagger_{C_1C_2C_3C_4} \hat{A}^\dagger_{D_1D_2D_3D_4}\\
&&\hspace{6em}\epsilon^{A_1B_1}\cdots\epsilon^{A_4B_4}\,\epsilon^{C_1D_1}\cdots\epsilon^{C_4D_4} |0\rangle\,.
\eena
We can now rearrange indices using identities for the $\epsilon$ tensors, such as
\ben
\epsilon^{A_1B_1}\epsilon^{C_1D_1}=\epsilon^{A_1C_1}\epsilon^{B_1D_1}-\epsilon^{A_1D_1}\epsilon^{B_1C_1}
\label{epsilons}
\een
which can be represented diagrammatically as
\begin{equation}
\includegraphics[scale=0.2,valign=c]{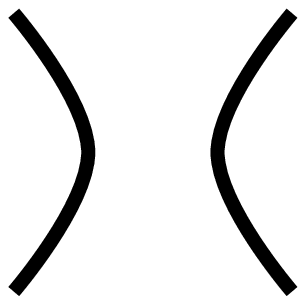}=-
\includegraphics[scale=0.2,valign=c]{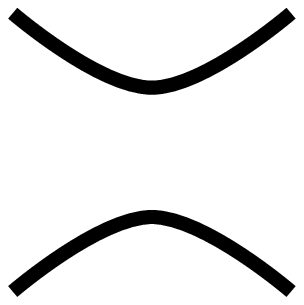}-\includegraphics[scale=0.2,valign=c]{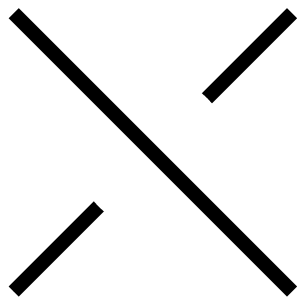}\,.
\end{equation}
Inserting (\ref{epsilons}) into (\ref{twodipoles}), we find that a state of two disconnected three-spheres would necessarily be equivalent to a sum of two states representing connected manifolds, 
\begin{equation}
\includegraphics[scale=0.2,valign=c]{dipole.eps}\,\includegraphics[scale=0.2,valign=c]{dipole.eps}\big|0\rangle=-\includegraphics[scale=0.2,valign=c]{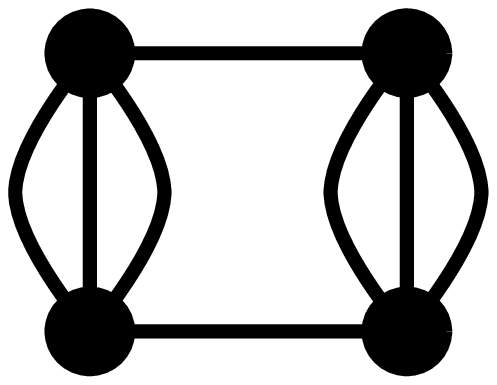}\big|0\rangle
-\includegraphics[scale=0.2,valign=c]{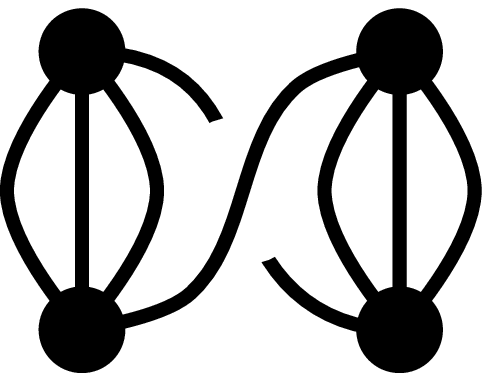}\big|0\rangle=-2\includegraphics[scale=0.2,valign=c]{connected.eps}\big|0\rangle\,.
\end{equation}
This argument obviously extends to more complicated states; any topological information extracted from Fock states in our model must hence come from elsewhere, not from a representation of their structure in terms of graphs. The fact that these Fock states, unlike LQG spin network states, cannot be associated unambiguously to graphs also applies to the full GFT setting, and is an important difference between the LQG and GFT Hilbert spaces \cite{LQGGFT}. 

\section{Discussion}
In this paper, we constructed a toy model for quantum cosmology in the framework of group field theory (GFT). The model realises two basic principles: that geometric observables have discrete eigenvalues in quantum gravity and that cosmological expansion is realised by creating new quanta rather than inflating existing ones. Since the cosmological evolution must change, therefore, the number of quanta, dynamics is best formulated in a second quantised framework, as given in group field theory. The model itself is formulated on the full GFT Fock space, only the dynamics differs from usual GFT models. We showed, in fact, that the cosmological expansion for an FLRW universe filled with a free massless scalar field can be modelled on the GFT Fock space by a squeezing operator. The resulting Schr\"odinger equation can be integrated trivially. Starting from a suitable initial state, such as the Fock vacuum, the expectation value of the total three-volume evolves according to modified Friedmann equations, which are very similar to those previously obtained in GFT cosmology \cite{GFTfried} and loop quantum cosmology \cite{LQC, improdyn}. The classical Big Bang singularity is then replaced by a quantum bounce connecting the contracting and expanding branches. 

The kinematics of our model is taken from both LQG and GFT. The configuration variables are given by $SU(2)$ holonomies along four distinct links meeting at a vertex, with each such vertex representing a tetrahedron. In the quantum theory, the volume of this tetrahedron can only assume certain discrete eigenvalues \cite{bianchihaggard}. The dynamics, on the other hand, were constructed without direct input from LQG or GFT. The starting point was the following observation: given a conventional Wheeler--DeWitt minisuperspace quantisation, the cosmological expansion would be generated by a dilatation operator $\sim \mathrm{i}V\partial_V$. But such an operator cannot exist on the GFT Fock space, because the differential $\partial_V$ is not well-defined if the volume has a discrete spectrum. We have to replace $\mathrm{i} V\partial_V$ by a finite difference operator, and we saw that a squeezing operator provides a particularly simple candidate for such an operator. By imposing the additional simplifying assumption that only isotropic tetrahedra are excited, and excited only in the fundamental $j=\tfrac{1}{2}$ representation of $SU(2)$, we reduced the field theoretic GFT formalism to a simple matrix model, defined in terms of oscillator modes $A^i(\phi)$ that only depend on the value of the massless scalar field $\phi$ used as time. Thus, inspired by both LQG and GFT, we constructed a certain \emph{matrix cosmology}.

In quantum gravity, matrix models have appeared in various contexts before, from quantum gravity in two dimensions \cite{ginsparg} to a possible non-perturbative definition of M-theory \cite{Mtheory}. Applications of such matrix models to quantum cosmology were discussed in, e.g.~\cite{Mcosmology}. In this context one has matrices $X^i(t)$ that represent spacetime coordinates at the quantum level. This is conceptually different from our more abstract \emph{background-independent} oscillators $A^i(\phi)$, but the basic objective for quantum cosmology is the same, namely to derive effective Friedmann equations for an effective scale factor $a(t)$ (in our work, such effective Friedmann equations are derived from the expectation values of the three-volume $V(\phi)$ as a function of $\phi$). The model that we developed here could inspire, therefore, further developments relating GFT cosmology to approaches of \emph{matrix cosmology} that come from other corners of quantum gravity research.

Our model shows explicitly how physical solutions to a many-particle quantum cosmology model can lead, in principle, to states of condensate type, as used previously in the context of GFT \cite{GFTreview}. The effective dynamics for such states reproduces then the main features of classical cosmology and LQC. Finally, we also commented on the impossibility of associating a unique spatial topology to our quantum states. This is a consequence of the chosen statistics: states that would be distinguishable in the LQG Hilbert space may be realised as the same quantum state in GFT.

As regards the fundamental definition of GFT models, the main new ingredient at the kinematical level was the use of equal-time commutation relations for the fundamental GFT field operators (and thus, for the oscillator mode operators). Such commutation relations have not been used in GFT before, but are suggested once we deparametrise the Wheeler\,--\,DeWitt equation with respect to a distinguished time variable (in our case this is the value of the scalar field $\phi$). Further work is needed to elucidate the precise relation of this formalism to the usual one in which no deparametrisation, and no equal-time commutator algebra, is used.

\section*{Acknowledgments}
This research was supported in part by Perimeter Institute for Theoretical Physics. Research at Perimeter Institute is supported by the Government of Canada through the Department of Innovation, Science and Economic Development Canada and by the Province of Ontario through the Ministry of Research, Innovation and Science. The work of SG was supported by a Royal Society University Research Fellowship (UF160622).


\end{document}